\documentclass[conference]{IEEEtran}
\usepackage{array}
\usepackage{mathptmx}      
\usepackage{amsmath}       
\usepackage{amsthm}
\usepackage{pbox}
\usepackage{verbatim}
\usepackage{color}
\usepackage{amssymb}
\usepackage{cite}
\usepackage[utf8]{inputenc}
\usepackage[english]{babel}

\usepackage{graphicx}
\graphicspath{{./img/}}
\DeclareGraphicsExtensions{.pdf}
\usepackage{enumerate}

\newcommand{\myincludegraphics}[2]{
\begin{figure}
  \centering
  \includegraphics[scale=0.65]{img_#1}
  \caption{#2}
  \label{fig:#1}
\end{figure}
}

\newcommand{\myincludegraphicstwo}[2]{
\begin{figure}
  \centering
  \includegraphics[scale=0.60]{img_#1}
  \caption{#2}
  \label{fig:#1}
\end{figure}
}

\newcommand{\myincludegraphicsthree}[2]{
\begin{figure}
  \centering
  \includegraphics[scale=0.31]{img_#1}
  \caption{#2}
  \label{fig:#1}
\end{figure}
}

\newcommand{\myincludegraphicsfour}[2]{
\begin{figure}
  \centering
  \includegraphics[scale=0.18]{img_#1}
  \caption{#2}
  \label{fig:#1}
\end{figure}
}

\newcommand{\myincludegraphicsfive}[2]{
\begin{figure}
  \centering
  \includegraphics[scale=0.44]{img_#1}
  \caption{#2}
  \label{fig:#1}
\end{figure}
}

\newcommand{\myfigref}[1]{Figure~\ref{fig:#1}}
\begin{document}
\title{Co-simulation for Cyber Security Analysis: Data Attacks against Energy Management System}

\author{\IEEEauthorblockN{Kaikai Pan\IEEEauthorrefmark{1},
Andr\'e Teixeira\IEEEauthorrefmark{2},
Claudio David L\'opez\IEEEauthorrefmark{1}
and Peter Palensky\IEEEauthorrefmark{1}} 
\IEEEauthorblockA{\IEEEauthorrefmark{1}Intelligent Electrical Power Grids\\
Faculty of EEMCS, Delft University of Technology,
Delft, The Netherlands}
\IEEEauthorblockA{\IEEEauthorrefmark{2}Engineering Systems and Services\\
Faculty of TPM, Delft University of Technology,
Delft, The Netherlands}}

\maketitle

\begin{abstract}
It is challenging to assess the vulnerability of a cyber-physical power system to data attacks. In order to support vulnerability assessment, with the exception of analytic methods, a suitable platform for security tests needs to be developed. In this paper we analyze the cyber security of energy management system (EMS) against data attacks. First we extend our analytic framework that characterizes data attacks as optimization problems with the objectives specified as security metrics and constraints corresponding to the communication network properties. Second, we build a platform in the form of co-simulation - coupling the power system simulator DIgSILENT PowerFactory with communication network simulator OMNeT++, and Matlab for EMS applications (state estimation, optimal power flow). Then the framework is used to conduct attack simulations on the co-simulation based platform for a power grid test case. The results indicate how vulnerable of EMS to data attacks and how co-simulation can help assess vulnerability.  

\end{abstract}

\IEEEpeerreviewmaketitle

\section{Introduction}

Cyber security vulnerabilities within the information and communication technology (ICT) infrastructure may allow attackers to manipulate the physical system, communication network or software applications in the cyber-physical power system. As a real example of cyber attack reported recently, highly destructive malware corrupted automation systems in substations resulting in a large scale blackout in the Ukrainian power grid \cite{Goodin2016}. Modern energy management systems (EMS) combined with Supervisory Control and Data Acquisition (SCADA) networks provide support for the monitoring and control of power grids. However, this critical infrastructure is vulnerable to cyber attacks and several attack events have been reported, see \cite{Gorman2009, ChenAbu-Nimeh2011}. In order to increase the security of these systems, one needs analytic methods to first understand the vulnerabilities and then to validate or explore them with appropriate tools. Some of the literature has already tackled these problems. Vulnerability assessment methods mainly using analytic expressions have been proposed in \cite{KosutJiaThomasEtAl2011, Hug2012, Pan2016}. Some tools based on co-simulation techniques to integrate simulated power systems, communication network and controls have been developed to analyze the behavior of cyber-physical power systems including cyber security issues \cite{Stifter2014, Wei2014, Mallouhi2011}.     

However, these two parts of the work are usually conducted independently even though they are related. Analytic methods may have to ignore some details when modeling the heterogeneous cyber-physical system, but could be used to guide the cyber security tests on co-simulation tools, while the tools can support the security analysis with empirical results. This could contribute to develop more robust algorithms/methods that combine system-theoretic and ICT-specific measures to protect EMS against data attacks \cite{Findrik2016}. In this paper, we aim to contribute in closing this gap by extending the typical vulnerability assessment framework to incorporate communication network properties and developing a co-simulation platform to conduct simulations on data attacks against EMS. In order to achieve this, some communication network properties are modeled in the analytic vulnerability assessment framework. Additionally, experiments are conducted on the developed co-simulation platform and the simulation results are analyzed.

The outline of the paper is as follows. Section II details the problem statement and our motivations on developing the methods and tools. In Section III, the analytic vulnerability assessment framework is illustrated. We further analyze what communication network properties should be considered in order to extend the framework. The co-simulation platform is presented in detail in Section IV, including how the power system and communication network are modeled, how the tools are integrated and how the attacks are implemented in OMNeT++. Section V shows the empirical results from co-simulation. We also provide a discussion on combining system-theoretic and ICT-specific measures to protect EMS. The conclusion remarks are in Section VI. 

\section{Problem Statement and Motivation} 

\subsection{Data Attacks Against Energy Management System}

The SCADA system supports the EMS of the information delivery as indicated in \myfigref{SEwithAttacks}. As a core part of EMS, State Estimation (SE) provides the operator an estimate of the state of the electric power system. SE uses measurements collected by the Remote Terminal Units (RTUs) in substations and transmitted through the SCADA communication network to the Master Terminal Units (MTUs) in the control center. The estimated state information is then processed by other applications in EMS such as optimal power flow (OPF) and Contingency Analysis (CA) to compute optimal control action while ensuring reliability and safety. The critical nature of EMS highlights the importance of making it accurate and secure for power grid operations.

\myincludegraphicstwo{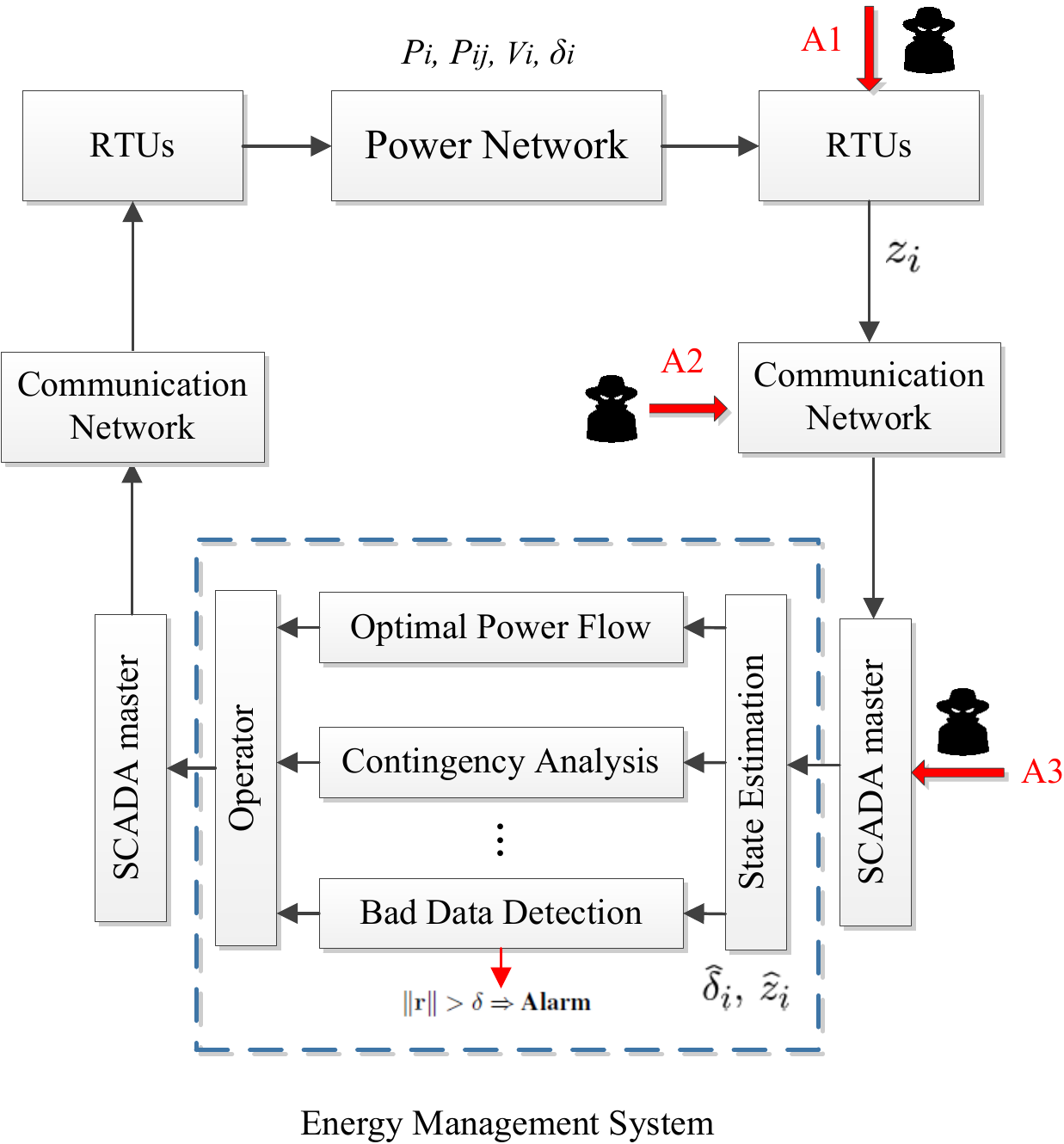}{A schematic block diagram of the power network, SCADA system and EMS. The SE uses power flow ($P_{i}$, $P_{ij}$) measurements ($z_{i}$) collected by RTUs and transmitted by the SCADA system to estimate the current state ($\delta_{i}$) of the power network. An alarm is triggered by the Bad Data Detection (BDD) when the norm of the measurement residual $r$ exceeds a given threshold. The cyber attack can manipulate the measurements by directly tampering the RTUs (A1), the SCADA communication network (A2) or even the SCADA master (A3). Figure adapted from \cite{Sandberg2010}.}

However, as SCADA systems become more connected to the Internet and corporate networks, they are potentially vulnerable to a large number of security threats. This is one motivation of our work. Substations need remote access connection for monitoring and maintenance, which may expose them to cyber attacks. Besides, for most industrial communication protocols, e.g., DNP 3.0, IEC 61850, adequate cyber security features were not always included at the time of publishing \cite{HongChenLiuEtAl2015}. As shown in \myfigref{SEwithAttacks}, the manipulation of measurements can arise from various levels (A1, A2, A3). 

\subsection{Towards Cyber-Secure and Resilient State Estimation}

Assuming that the power system has $n+1$ buses, the typical state estimation technique solves the following problem under DC power flow model,
\begin{equation}
\label{eq: AC_SE}
z = Hx + e,
\end{equation}
where the vector $z$ denotes the $m$ measurements, $H \in \mathbb{R}^{m \times n}$ represents the system 
model that describes the dependencies of measurements and system state, $x \in \mathbb{R}^{n}$ is the state vector of $n$ bus phase angles except the reference one, $e$ is the measurement noise vector which is always assumed to have a Gaussian distribution of zero mean and covariance matrix $R = \mbox{diag}(\sigma_{1}^{2}, \ldots, \sigma_{m}^{2})$. For such a large-scale SCADA system, lost data, inaccurate measurements and failing RTUs or other infrastructures in communication network are common \cite{Sandberg2010}. Thus there is a built-in Bad Data Detection (BDD) scheme to deal with that. In BDD, the residual signal $r$ is evaluated to detect and locate existing anomalies of data, as depicted in \myfigref{SEwithAttacks}. However, such kind of system-theoretic measure is not adequate to protect the EMS against potential data attacks. The data can be corrupted in a coordinated way that still fulfills the power flow laws and would not be detected by BDD \cite{LiuNingReiter2009}. 

A considerable amount of work has been done on vulnerability assessment of data attacks against EMS \cite{KosutJiaThomasEtAl2011, Hug2012, Pan2016, Rahman2013, VukovicSouDanEtAl2012}. Usually these are system-theoretic measures that are based on analytic methods. Another group of measures from ICT-specific security includes firewalls, network intrusion detection systems and authentication, etc. Recently some organizations (e.g. NIST, NERC) have proposed security standards that combine the measures from ICT-specific and system-theoretic ones \cite{Findrik2016}. Regarding these issues, we have the following recommendations:
\begin{itemize}
\item The system-theoretic measures based on analytic methods need empirical results for validation and analysis;
\item The vulnerability assessment of data attacks should take the attack impact/consequences into account;
\item To improve the security of EMS, there is a necessity to explore the interactions between system-theoretic and ICT-specific measures and try to combine them. 
\end{itemize}

To support the security analysis above, an integrated platfrom using various tools including simulators for power network, SCADA communication network and EMS applications could offer these capabilities. Co-simulation is currently one of the most popular methods to analyze such a large, heterogeneous cyber-physical system \cite{Palensky_2017}. Therefore in this paper we propose to extend our current analytic vulnerability assessment methods to incorporate communication network properties and enable them with support from a co-simulation platform.   

\section{Analytic Framework Incorporating Communication Network Properties}

\subsection{Data Attacks and Vulnerability Assessment Problem}

With the goal of perturbing the SE and further corrupting the applications in EMS, the attacker would gain access to the measurement data through various levels (A1, A2, A3) as shown in \myfigref{SEwithAttacks}. The measurements under different attack scenarios from the view of SE can be presented as follows:
\begin{itemize}
\item Data integrity attack - also known as false data injection (FDI) attack, is able to change measurements values from $z$ to $z+a$ where $a$ is the \emph{FDI attack vector}.

\item Data availability attack - includes DoS or jamming attack which would make specific measurements unavailable to SE, i.e., $z_{0} = (I - \mbox{diag}(d))z$ where $d \in \{0,1\}^{m}$ is the \emph{availability attack vector} and $I$ is an identity matrix.

\item Combined attack - combines the FDI and availability attack that makes the measurements from $z$ to $(I - \mbox{diag}(d))z + a$ corrupted by $a$ and $d$. 
\end{itemize}

The vulnerability assessment is presented through the notion of \emph{security metric} which computes how many measurements need to be manipulated by the attacker to keep stealth against the BDD. This metric can quantify the attack resources and consequently the vulnerabilities of EMS to attacks. The EMS is more vulnerable to attacks with small security metric since such attack needs less resources to be executed. Taking the attack scenarios under DC model as an example, if the attacker corrupts certain measurements using FDI attack vector $a = Hc$, 
it can remain hidden from the BDD but perturb the current state to a degree of $c$ \cite{LiuNingReiter2009}. It's also shown in our recent work \cite{Pan2016} that combined attacks can achieve the same target with the attack vector $a = (I - \mbox{diag}(d))Hc$. It should be noted that these data attacks are assumed not to make the system unobservable. 
In sight of this, it is natural to consider the following security metric problem:
\begin{subequations}\label{sec_idx_com_o}
\begin{align}
\alpha_{j}:= \min\limits_{c,d}\quad& \lVert a \rVert_{0} + \lVert d \rVert_{0} \nonumber\\
\mbox{s.t.}\quad& a = H_{0}c, \label{eq:2a}\\
 & H_{0} =(I - \mbox{diag}(d))H, \label{eq:2b}\\ 
 & a(j) = \mu, \label{eq:2c}\\
 & d(i) \in \{0,1\}\quad\mbox{for all } i, \nonumber
\end{align}
\end{subequations}
where $\lVert a \rVert_{0}$ and $\lVert d \rVert_{0}$ denote the number of non-zero element in the vectors. Here $\mu$ is a non-zero value denoting the attack magnitude on measurement $j$, and $\alpha_{j}$ is the security metric that can illustrate how many measurements or RTUs needed by the attacker to corrupt EMS and keep stealth.  

\subsection{Analytic Vulnerability Assessment Incorporating Communication Network Properties}

The vulnerability assessment problem in \eqref{sec_idx_com_o} suits for the cases that attacks arise from the level of A1 in \myfigref{SEwithAttacks}. This security metric directly shows that manipulation on several RTUs is needed for the attacker. However in practice, tampering with RTUs directly becomes much harder as more RTUs are authenticated and secured. A more interesting scenario is to look into attacks from the level of A2 since usually attacks would explore vulnerabilities in communication networks, e.g., compromising remote access points, obtaining access to corporate networks. The vulnerability assessment should consider the communication network. However, modeling the communication network in an analytic framework is challenging due to its complexity and heterogeneity. Here, the communication network properties of interest for security analysis are as follows:
\begin{itemize}
\item Communication topology;
\item Routing schemes - the routing paths of packets / data;
\item Communication latency - how the packets / data would be delayed in each communication infrastructure;
\item Packet loss / missing data - the possibility of packet drop in each communication infrastructure.
\end{itemize}   

Here we introduce a method to deal with the first two properties that can be employed in the analytic vulnerability assessment. Another two properties of communication networks, latency and packet loss, could also be incorporated into analytic framework, not for vulnerability assessment but for combining ICT-specific measures and system-theoretic measures. We show such potentials in Section V. Let us consider a simple communication network as shown in \myfigref{Routingpath}. We can describe it as an undirected graph $\mathsf{G} = (\mathsf{V}, \mathsf{E})$ where $\mathsf{V}$ is the set of nodes and $\mathsf{E}$ is the set of communication links. Assuming that a measurement $i$ would be transmitted through a routing path $P1$, we establish a binary vector called \emph{routing vector},
\begin{equation}
\label{eq: routingvector}
r_{i,P1} = [r_{vi,P1}^{T},r_{ei,P1}^{T}],
\end{equation}
where in routing vector $r_{vi,P1} \in \{0, 1\}^{N}$ denotes the part corresponding to nodes and the entries are equal to 1 if the route traverses the node. $r_{ei,P1} \in \{0, 1\}^{E}$ denotes the part corresponding to communication links and the entries are equal to 1 if the route traverses the link. $N$ and $E$ denote the whole number of nodes and edges in the graph. Thus for the path $P1$, we can obtain
\begin{equation}
\label{eq: routingvector_real}
r_{vi,P1} = [1, 1, 0, 1]^{T}, r_{ei,P1} = [1, 0, 1]^{T}.
\end{equation}

\myincludegraphicstwo{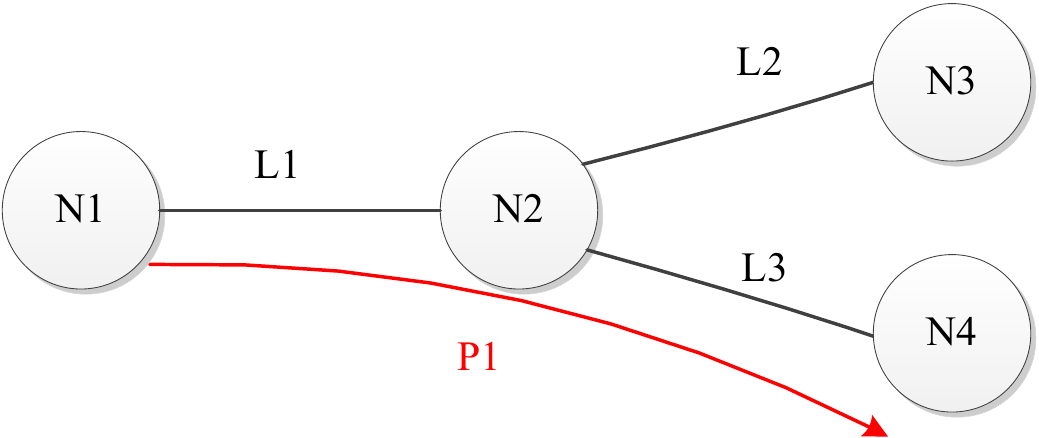}{A simple communication network for illustration of routing path. $N1, N2, N3, N4$ represent communication nodes and $L1, L2, L3$ represent communication links. The routing path $P1$ follows $N1-L1-N2-L3-N4$.} 

Using the graph of the communication network and routing schemes for all the measurements, we can build a \emph{routing matrix} and each row of the matrix is a \emph{routing vector}. The \emph{routing matrix} and \emph{routing vectors} contain the information of communication topology and routing schemes. In our recent work \cite{Pan2016}, we extend the vulnerability assessment problem \eqref{sec_idx_com_o} to the following one,
\begin{subequations}\label{sec_idx_sub_withD}
\begin{align}
\beta_{j}:= \min\limits_{c, d, x, y} \quad & \lVert x \rVert_{0} + \lVert y \rVert_{0} \nonumber \\
\mbox{s.t.} \quad& a = H_{0}c, \\
 & H_{0} =(I-\mbox{diag}(d))H, \\ 
 & a(j) = \mu, \\
 & a(i) = 0 \mbox{ if } r_{vi,P} = 0,\mbox{ for all } i \not= j, P, \label{eq:5d} \\  
 & d(i)\leq r_{vi,P}x + r_{ei,P}y \quad\mbox{for all } i \not= j, P, \label{eq:5e} \\
 & d, x, y \quad \mbox{are all binary vectors}, \nonumber
\end{align}
\end{subequations}
where $x \in \{0, 1\}^{N}$ and $y \in \{0, 1\}^{E}$ are vectors whose entries are 1 if certain nodes/links are attacked. The constraints \eqref{eq:5d} and \eqref{eq:5e} use the \emph{routing vectors} to map the data attacks on measurements to attacks on communication network. They also indicate that for FDI attack on measurement $j$, at least one node should be attacked and included on all of its routing paths and for availability attack on measurement $j$, at least one node or communication link should be attacked and included on all of its routing paths. This is the worst-case scenario that the attacker is assumed to have the knowledge of both communication network (topology and routing schemes) and power system network (the network model $H$). The metric $\beta_{j}$ can illustrate the vulnerability of EMS to data attacks on the communication network. It should be noted that some ICT-specific security measures can be modeled in \eqref{sec_idx_sub_withD}. For instance, multi-path routing schemes can be described using \emph{routing vectors} in constraints \eqref{eq:5d} and \eqref{eq:5e}. Data authentication can be implemented by adding constraints to indicate which measurement originates from the node with authentication is protected. 

These two analytic vulnerability assessment problems \eqref{sec_idx_com_o} and \eqref{sec_idx_sub_withD} can be formulated as mixed integer linear programming (MILP) problems. Further details on formulations and solutions can be found in \cite{Pan2016}. However, these security metrics do not consider the attack impact on the physical system operation. In fact, data attacks with the same security metrics could have considerable different impact. Co-simulation could offer the capabilities to look into the attack impact and provide empirical results to validate and contribute in developing mitigation measures, as discussed in Section II-B.         

\section{Co-simulation Supporting Vulnerability Assessment and Security Analysis}

\subsection{Co-simulation Tools}

An integrated environment including simulators of power system, communication network and EMS applications is needed for security analysis. In order to allow for real-time analysis of cyber attacks, the co-simulation platform is implemented with three tools: DIgSILENT PowerFactory for the power system, OMNeT++ for the communication network, and Matlab/Matpower for the EMS algorithms. They are coupled as shown in \myfigref{cosimdiagram}. Here, measurements of the power flow going in and out of each bus of the power system simulated in PowerFactory are sent to the EMS applications in Matlab through a communication network simulated in OMNeT++. 

\myincludegraphics{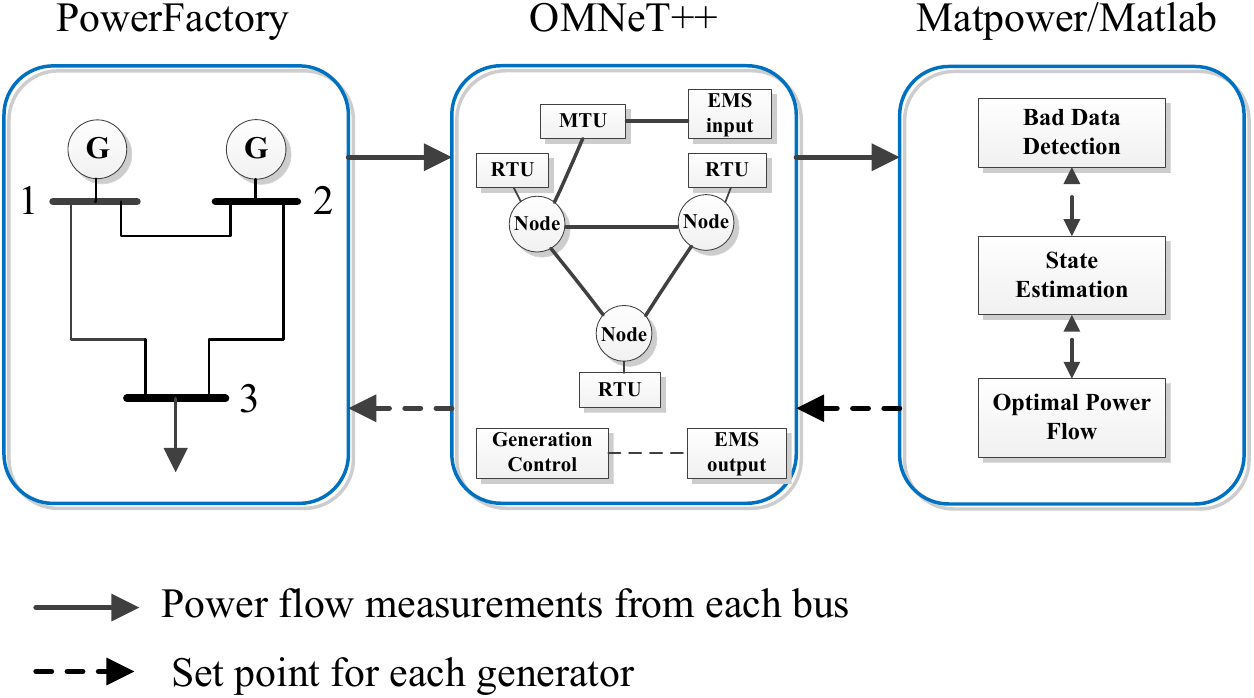}{Co-simulation diagram.}

\subsubsection{Power system simulator} DIgSILENT PowerFactory is used to conduct a quasi-static power flow simulation. PowerFactory's Python API is used to create a script that controls the execution of the simulation. The same script implements the interface with OMNeT++. Real time execution is achieved by synchronizing the power flow calculations with the system clock. The script sends measurements to OMNeT++ every fixed time (set to be 5 seconds), but it can expect generator set points at any time. Thus, a dedicated thread that received set points and sets them in the power system model is required. This thread sets the generators according to the set points as soon as they arrive, unless a power flow calculation is being executed, in which case it waits for the calculation to finish.

\subsubsection{Communication Network Simulator} OMNeT++ is used for discrete-event based communication network simulation. The communication model in OMNeT++ is shown in \myfigref{scadanetwork}. A custom OMNeT++ \emph{scheduler} is built to enable data exchange with PowerFactory and Matlab over TCP/IP sockets and run the OMNeT in real-time. In \myfigref{scadanetwork}, \emph{RTU} is a module served by the scheduler and acts as a RTU proxy. The second module developed called \emph{MTU} works as master unit and data concentrator that receives packets and has a FIFO queue. There is a \emph{Modem} module that acts as a communication bridge and a \emph{Router} module with routing table for the packets. Thus, the RTU, Modem and Router represent the LAN (local area network) of a substation. Besides, the module \emph{EMSInput} and \emph{EMSInout} provide measurements to EMS and receive set points from EMS in Matlab respectively. For the message implementation, a new packet class \emph{MeasurePacket} is derived to contain the measurement data and be used by all the modules and scheduler. There are two kinds of communication channels: channel of the LAN and channel of the WAN (wide area network) between routers. Different latency and packet loss probability parameters are set in these two channels. It should be noted that implementation of a real SCADA system with protocols (e.g., IEC61850, DNP3.0) and hierarchical network structure that is close to reality in OMNeT++ is not our focus in this paper. Instead we try to explore how co-simulation can support the analytic vulnerability assessment. 

\myincludegraphicsthree{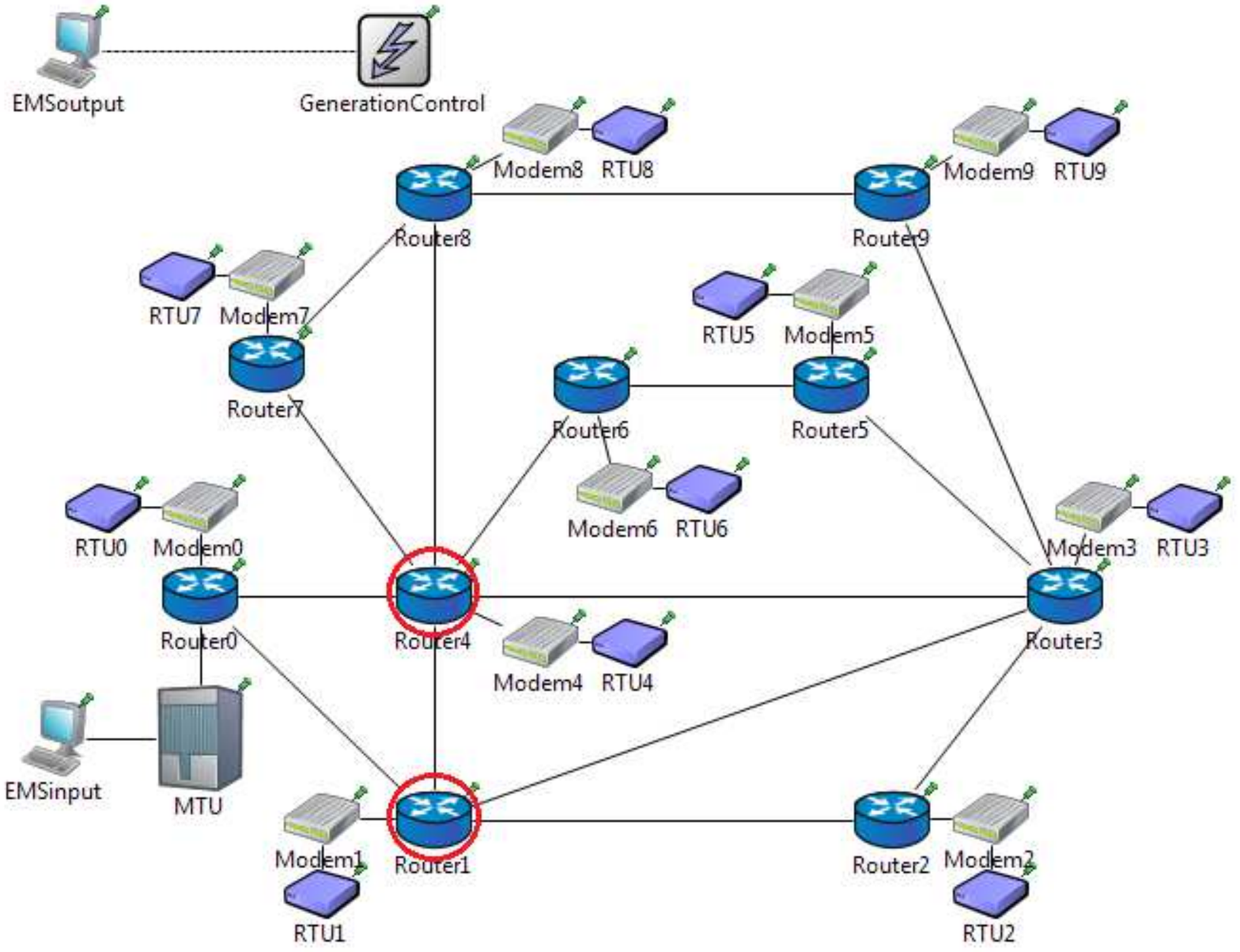}{Test communication network of 14 bus system in OMNeT++.}      

\subsubsection{EMS algorithm} Matpower has been used to simulate the EMS applications in Matlab, including state estimation (with bad data detection) and optimal power flow algorithms. A script is implemented to exchange data with OMNeT++ scheduler over TCP/IP sockets and store measurements into a data pool. The State Estimation module uses the latest measurements from data pool to create a snapshot of estimated power flow. For every fixed time (set to be 30 seconds), the Optimal Power Flow module uses load estimates from State Estimation to perform optimal power flow calculation (also see \myfigref{SEwithAttacks}) and sends commands of generator set points to PowerFactory through OMNeT++.

\subsection{Simulation Integration}

Data is exchanged between PowerFactory, OMNeT++ and Matlab via TCP/IP sockets using the ASN.1 protocol. On the PowerFactory side, this is implemented in the Python script that controls the simulator execution, while on the OMNeT++ side, this is implemented through a custom scheduler which adapted part of the work from \cite{Stifter2014}. This scheduler act as the ``master'' to coordinate the co-simulation, handle the data exchanges with PowerFactory and Matlab, and also run the OMNeT++ in a real-time mode. For the synchronization, all simulators would be started from a command after initialization and tagged with time stamps with the system clock.   

\subsection{Modeling attacks in OMNeT++} 

An attacker can manipulate the measurements by injecting false data, making it unavailable or both. After accessing a router, the attacker can launch data integrity and availability attacks on all the data traveling through it by executing a \emph{man-in-the-middle attack}. By jamming, DoS or physical attack, the attacker can block measurements in communication links. In this paper, we consider the worst case scenario that the attacker is intelligent enough with full knowledge of both the power system and communication network. The attacker would use the combined attack policy in \eqref{sec_idx_sub_withD}, i.e., try to remain hidden from the BDD and manipulate the minimum number of routers. Then the corrupted measurement vector becomes
\begin{equation}
\label{eq:attack2}
z_{a} = (I - \mbox{diag}(d)) z + a,
\end{equation}
where $a = (I - \mbox{diag}(d))Hc$ and $d$ denote the FDI attack and availability attack respectively. The results from the analytic work in \eqref{sec_idx_sub_withD} is used to choose the routers to be attacked. These attacks is implemented in OMNeT++ by changing the behavior of the router in case it is accessed by the attacker. 

\section{Simulation Results and Discussion}

We consider the IEEE 14 bus system in \myfigref{14bus} to perform the security analysis. Mapping with \myfigref{14bus}, the communication network as depicted in \myfigref{scadanetwork} is used. The modeling of the communication network of IEEE 14 bus system is adapted from \cite{VellaithuraiBiswasLiuEtAl2015}. There are ten substations (each circle represents a substation in \myfigref{14bus}) and the control center with MTU and EMS is located at the reference bus (i.e., Bus 1). There is an RTU, a modem and a router in each substation. The packets containing the measurements data would be routed through multiple routers before reaching MTU. We use the single-path routing scheme for each measurement. 

\myincludegraphicsfour{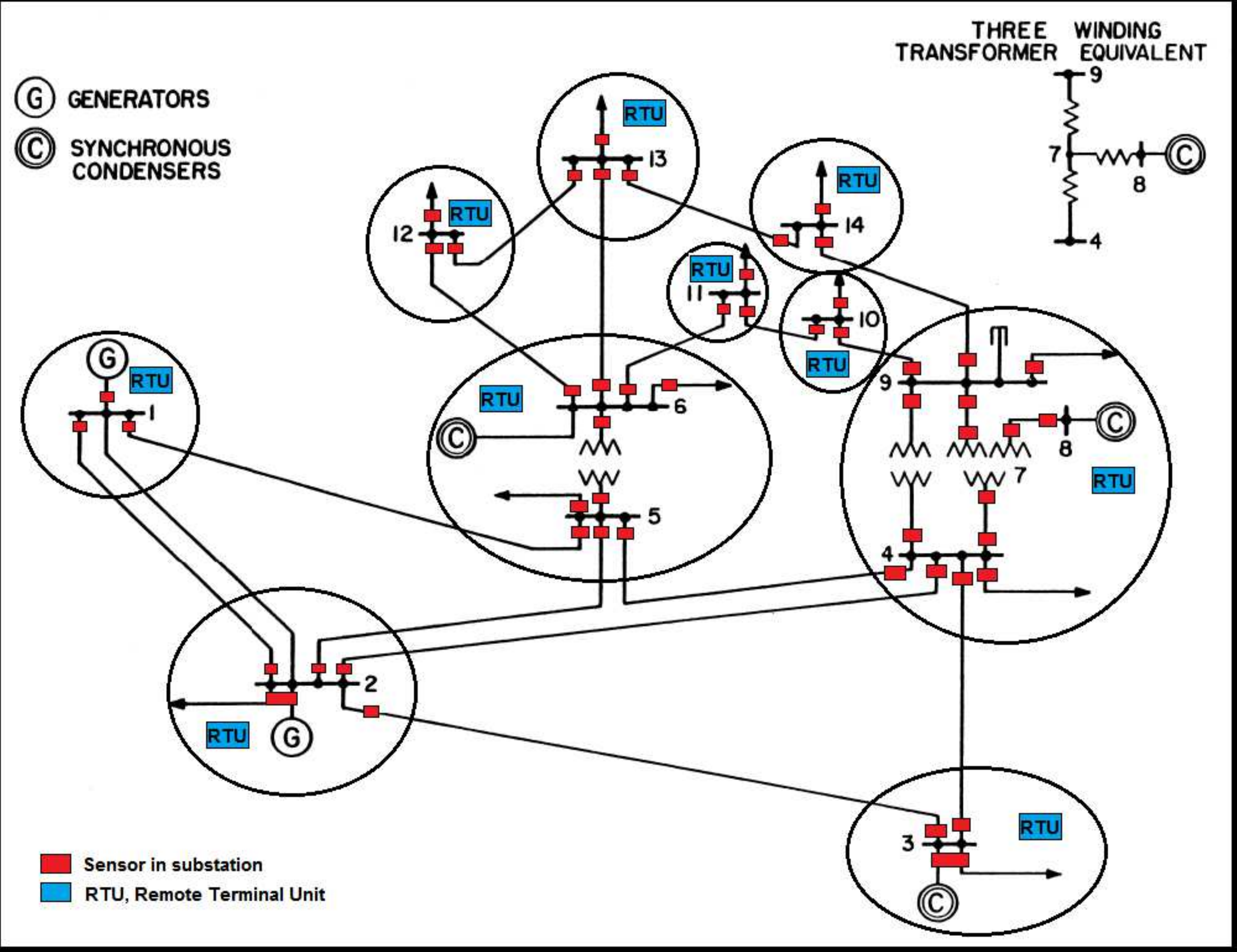}{IEEE 14 bus system. There are 2 generators. Bus 1 with Generator 1 is the reference/slack bus. Generator 2 is in Bus 2. The power flow measurements are collected in each bus and both sides of the branch.}  


The case of combined integrity and availability attack in Section IV-C has been implemented. The analytic results of \eqref{sec_idx_sub_withD} can be found in \cite{Pan2016}. It shows the minimum number of routers and links to be attacked in order to corrupt specific measurements and keep stealth. According to the analytic results, Router 4 (the backbone router) and Router 1 (marked with a red circle) are the most vulnerable network components. Thus we change the behavior of Router 4 and Router 1 independently to simulate the attack scenarios once an attacker gains access to their internals and the packets traveling through it. \myfigref{generation} shows the attack impact on the generation profile of generators in Bus 1 and Bus 2. \myfigref{powerflow} shows the attack impact on the active power flows when Router 1 is attacked.   

\myincludegraphicstwo{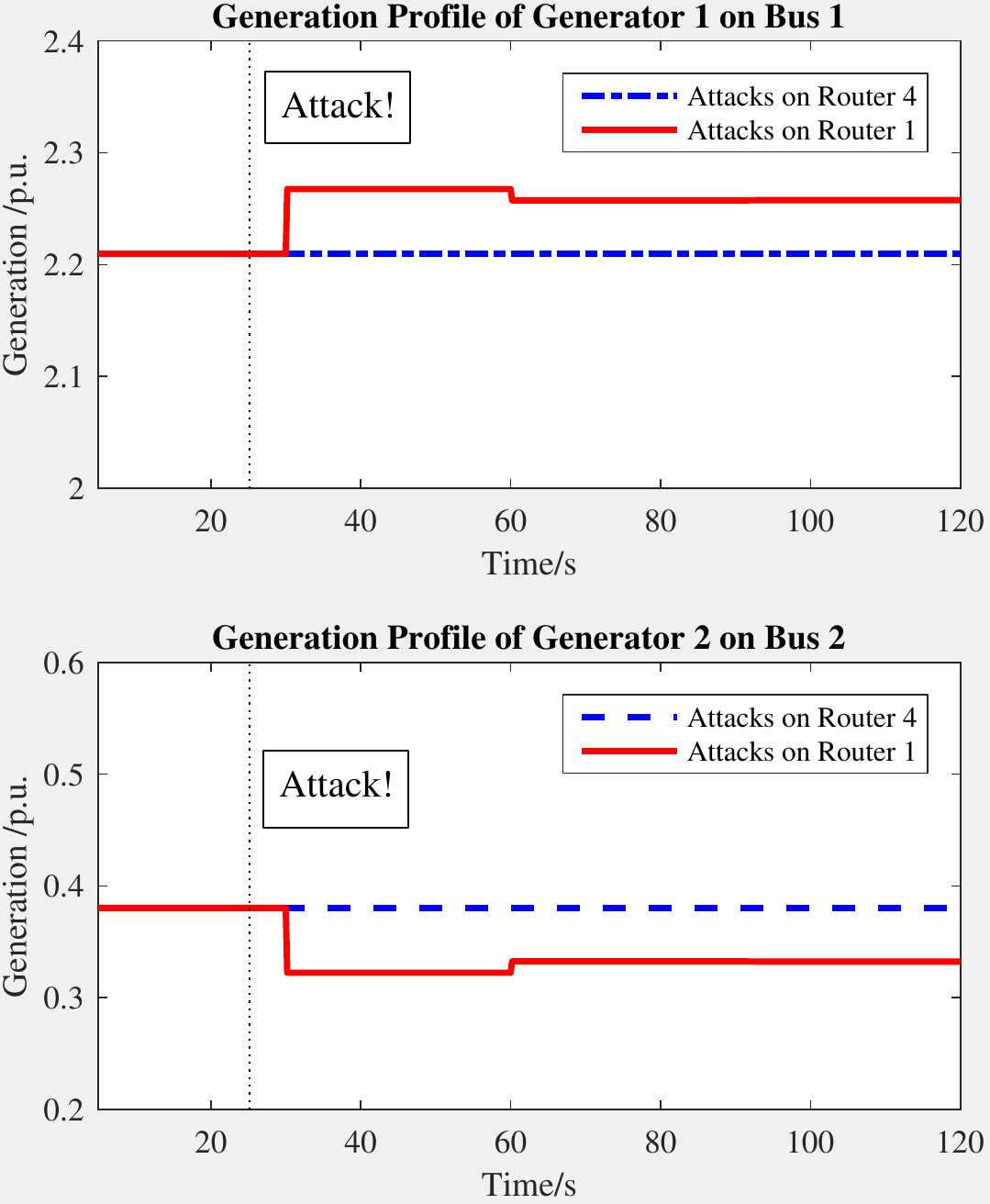}{Attack impact of stealth attacks on generation profile of Generator 1 and 2. The per-unit system is used and the power base is $100 MW$. The true power flow measurements are generated by DC power flow model with Gaussian noise ($\sigma_i = 0.005$ for all the measurements). Before the attack occurrence, the system is operating under the optimal power flow status giving the loads. In these two cases, the same number of measurements are corrupted.} 

\myincludegraphicsfive{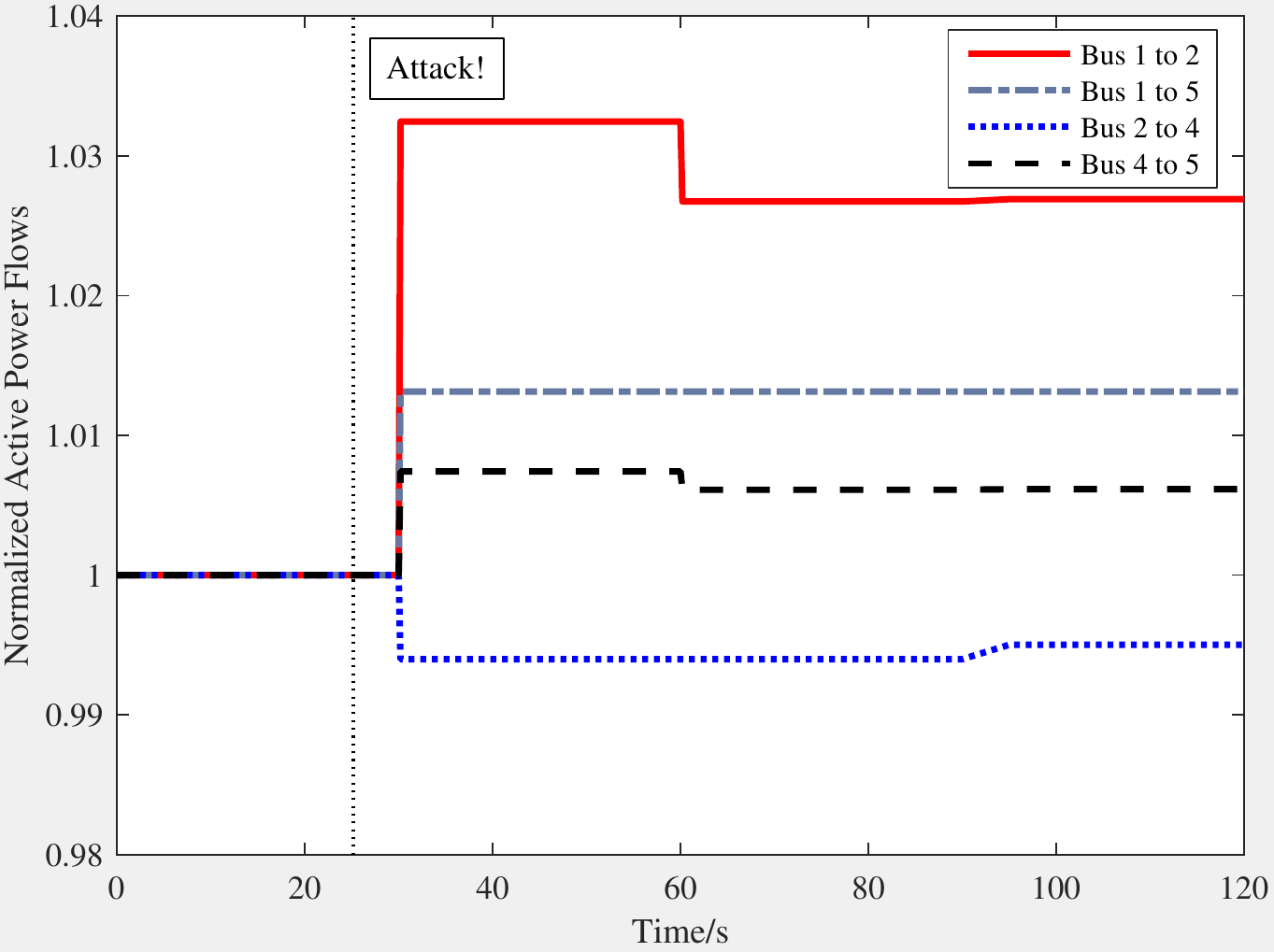}{Attack impact of stealth attacks on active power flows in the lines of bus 1 to 2, bus 1 to 5, bus 2 to 4, and bus 4 to 5. Router 1 is manipulated. The active power flows are normalized to the ones before the attack occurrence.}      

As shown in \myfigref{generation}, when Router 1 is attacked, the system ``fakes'' that the generation profile changes according to the set points. The generation of Generator 2 has decreased and Generator 1 should compensate. The ``latency'' between the attack occurrence and the change of generation profile is due to that the EMS sends out set points every 30 seconds. After the attack occurs, the generation profiles remain almost the same although the attack continues, which means the attack impact mainly depends on the initial attack magnitudes and measurements that are corrupted. When Router 4 is attacked, however, it seems that there is no attack impact on the generation profile, though Router 4 is the backbone router with the most number of packets traveling through. This is mainly because of the packets in or traveling through these two routers containing different measurements. According to our single path routing scheme, in Router 1, the attacker can gain access to the power flow  measurements on bus 2, 3 and 4, which has the major impact on the generation profiles of these two generators. For the case that Router 1 is attacked, the active power flows on the lines close to the generators are shown in \myfigref{powerflow}. The power flows get changed after redispatch according to the corrupted set points. Such physical impact can be utilized by the attacker to cause line overflows.      

\noindent\emph{Discussion on Combining Theoretic and ICT-specific Measures}

The proposed analytic vulnerability assessment method can be used to narrow down the attack scenarios. Using the co-simulation platform, the attack impact can be explored by directly simulating attacks. New security metrics could be formulated taking into account the impact of the data attack.

As discussed in Section II-B, co-simulation supports security analysis in combining the system-theoretic and ICT-specific measures. In the case of data attacks against EMS, the BDD scheme acts as a theoretic measure to detect bad data. However, it fails to trigger alarms when we simulate attacks on Router 1 and Router 4 since the measurements still fulfill the physical laws. To make it robust against data attacks, the communication network properties supported by co-simulation show the potentials for developing an advanced BDD scheme. For instance, when FDI attacks take place, the latency of attacked packets changes due to the attack process. When availability attacks occur, the latency of attacked packets can be treated as an extreme case. Thus a robust BDD scheme could be developed against combined attacks, incorporating network properties with the latency of packets measured in the co-simulation platform. We leave this for future work.      

\section{Conclusion}

In this paper, we contribute to extend analytic methods incorporating communication network properties and develop a co-simulation platform to analyze data attacks against EMS. The results shows the need to consider the vulnerability and attack impact in an integrated assessment framework and combine the system theoretic and ICT-specific measures to protect EMS. Our future work includes more security analysis on AC power flow model and other EMS applications using the co-simulation platform, developing robust algorithms for detection and mitigation measures, etc.     


\bibliographystyle{IEEEtran}
\bibliography{IEEEabrv,Literature}
%

\end{document}